\documentclass[RNAAS]{aastex631}

\usepackage{verbatim}

\accepted{\today}

\shorttitle{Evidence of early-stage tidal structures of open clusters}
\shortauthors{Li, Pang \& Tang}

\graphicspath{{./}{figures/}}

\begin{document}

\title{Evidence of Early-stage Tidal Structures of Open Clusters Revealed by Kinematics with Gaia EDR3 }

\author[0000-0002-5740-5477]{Yezhang Li}
    \affiliation{Department of Physics, Xi'an Jiaotong-Liverpool University, 111 Ren’ai Road, 
                Dushu Lake Science and Education Innovation District, Suzhou 215123, Jiangsu Province, P. R. China; Xiaoying.Pang@xjtlu.edu.cn.}

\author[0000-0003-3389-2263]{Xiaoying Pang}
    \affiliation{Department of Physics, Xi'an Jiaotong-Liverpool University, 111 Ren’ai Road, 
                Dushu Lake Science and Education Innovation District, Suzhou 215123, Jiangsu Province, P. R. China; Xiaoying.Pang@xjtlu.edu.cn.}

    \affiliation{Shanghai Key Laboratory for Astrophysics, Shanghai Normal University, 
                100 Guilin Road, Shanghai 200234, P. R. China}

\author[0000-0003-4247-1401]{Shih-Yun Tang}
    \affiliation{Lowell Observatory, 1400 W. Mars Hill Road, Flagstaff, AZ 86001, USA}
    \affiliation{Department of Astronomy and Planetary Sciences, Northern Arizona University, Flagstaff, AZ 86011, USA}

\newcommand{\pmra}{$\mu_\alpha \cos\delta$}
\newcommand{\pmdec}{$\mu_\delta$}
\newcommand{\masyr}{mas$\,$yr$^{-1}$}
\newcommand{\kms}{km\,$s$^{-1}$}

\begin{abstract}
Blanco 1, a 100\,Myr open cluster in the solar neighborhood, is well known for its two 50\,pc-long tidal tails. Taking Blanco 1 as a reference, we find evidence of early-stage tidal disruption in two other open clusters of $\sim$120\,Myr: the Pleiades and NGC\,2516, via Gaia EDR3 data.
These two clusters have a total mass of $2-6$ times that of Blanco\,1. Despite having a similar age as Blanco\,1, the Pleiades and NGC\,2516 have a larger fraction of their members bound: 86\% of their mass is inside the tidal radius, versus 63\% for Blanco\,1. 
However, a correlation between Blanco\,1's 50\,pc-long tidal tails and the ``kinematic tails'' in velocity space is also found for the Pleiades and NGC\,2516. This evidence supports the idea that the modest elongation seen in the spatial distribution for the Pleiades and NGC\,2516 is a result of early-stage tidal disruption.
\end{abstract}

\section{Motivation} \label{sec:intro}

\begin{figure}[t]
\label{fig:3d}
\centering
\includegraphics[height=19cm, width=18cm]{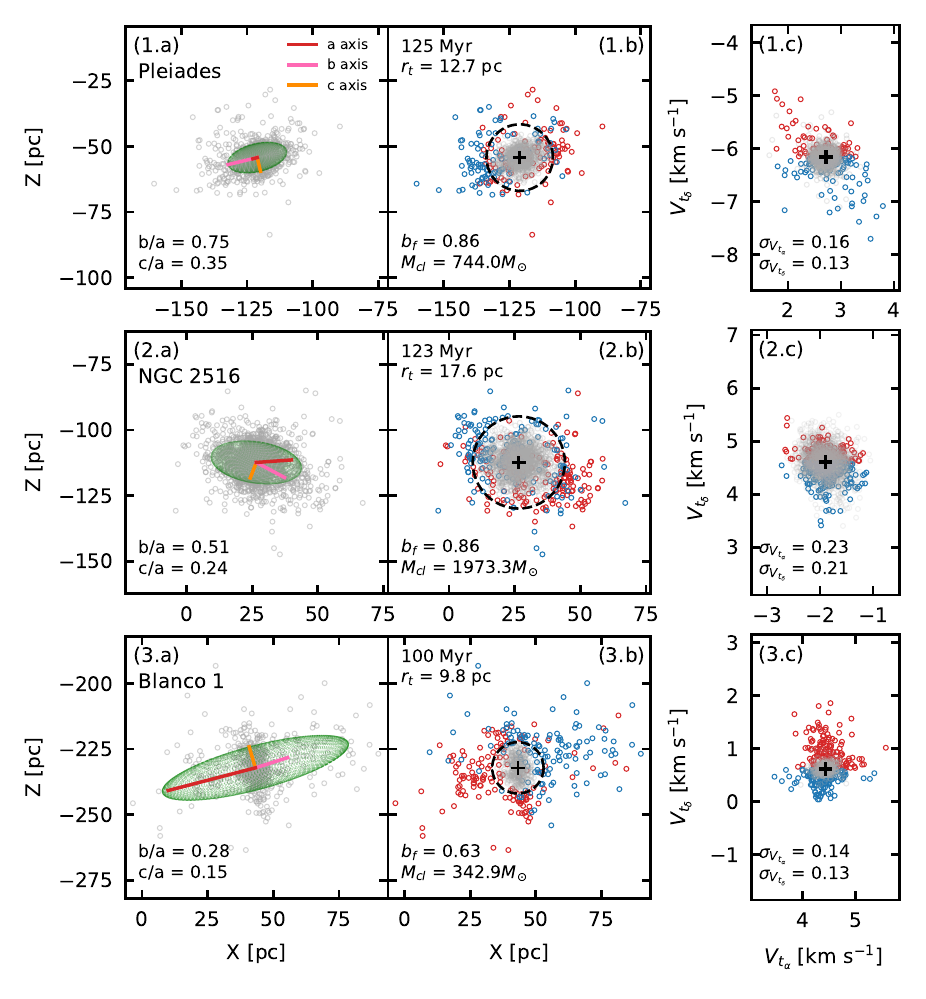}
    \caption{
    The spatial (a--b) and the tangential velocity (c) distributions of the Pleiades, NGC\,2516 and Blanco\,1. 
    The green ellipses in panels (a) represent the projection of the fitted ellipsoids onto the $X$-$Z$ plane. The three ellipsoid axes ($a$, $b$, and $c$) are indicated as red, pink, and orange. 
    The cluster centers are marked with an ($+$) in panels (b). Dashed black circles indicate the tidal radii. Grey circles in panels (b) and (c) represent stars located within tidal radius. $r_{\rm t}$ is the tidal radius, $b_{\rm f}$ the bound mass fraction, and $M_{\rm cl}$ the total mass. 
    The ($+$) symbol in panels (c) indicates the median value of the tangential velocity $V_{t_{\alpha}}$ and $V_{t_{\delta}}$, obtained from proper motions. The red and blue open circles in panels (b) and (c) respectively represent stars outside the tidal radius with $V_{t_{\delta}}$ values larger, or less than, the median $V_{t_{\delta}}$. The velocity dispersion is indicated in the bottom-left corner of panels (c). 
    }
\end{figure}

Spatially-elongated structures outside open clusters' tidal radius have been found in local samples of all ages \citep{pang2021}. However, extended tails $\gtrapprox$50\,pc are only discovered in star clusters older than $\sim$600\,Myr in the solar neighborhood \citep[e.g., Coma Berenice, Hyades, Praesepe:][]{tan19,roser2019a,roser2019b}, except for Blanco\,1 with 100\,Myr \citep{zhang20,pang2021}. Two other clusters, Pleiades and NGC\,2516 ($\sim$120\,Myr), with similar ages as Blanco\,1, have structures outside the tidal radius \citep{Lodieu19,pang2021}. Whether the small number of stars outside the tidal radius in the Pleiades and NGC\,2516 provides evidence of early-stage tidal tail formation remains unclear.
We therefore investigate the morphology of the Pleiades, NGC\,2516, and Blanco\,1, to search for the kinematic evidence supporting the nature of the tidal structures in clusters without significant elongation.

\section{Data analysis} \label{sec:analysis}

Cluster members used in this study are obtained from \citet[][NGC\,2516 with 2690 members and  Blanco\,1 with 703 members]{pang2021}, except for the Pleiades, which is not included in \citet{pang2021}.  We follow the same procedure in \citet{pang2021} to identify members of the Pleiades via the machine learning algorithm \texttt{StarGO} \citep{yuan18} with the latest Gaia Early Data Release 3 \citep[EDR 3;][]{gaia21}. We find 1407 member stars for the Pleiades, with a mean position of $(X, Y, Z) = (-121.2, +29.1, -54.2)$\,pc (in Heliocentric Cartesian coordinates), and mean proper motions of $(\mu_{\alpha} \cos\delta, \mu_{\delta}) = (+19.908, -45.341)$\,mas\,yr$^{-1}$. The member list for the Pleiades is  provided online\footnote{\dataset[Zenodo: 10.5281/zenodo.5067751]{https://doi.org/10.5281/zenodo.5067751}}.  Our identification of members is purely based on 5D phase-space information, without pre-assumptions about age, cluster size, and density threshold.

\section{Tidal Structures versus Kinematic ``Tails''}

We calculate the Heliocentric ($X,Y,Z$) coordinates  for the Pleiades, NGC\,2516, and Blanco\,1 using distances that have been corrected from the parallax errors \citep[see Section 4.1 in][]{pang2021}. Considering a distance range of 135.9--410.5\,pc for these clusters, the uncertainty in the corrected distance is about 0.8--2.6\,pc, according to the simulations of \citet[their Figure 4]{pang2021}.

To quantify the morphology of the clusters, we perform an ellipsoid fitting to their 3D spatial distribution \citep[see Section~4.3 in][]{pang2021}. Figure~\ref{fig:3d} panels (1.a--3.a) present the best-fitted ellipsoids (green ellipses) projected onto the $X$-$Z$ plane. The semi-major axis ``$a$'' represents the direction of morphological elongation, and the axes ratios $b/a$ (semi-intermediate/semi-major) and $c/a$ (semi-minor/semi-major) indicate the significance of the elongation. Although both the Pleiades and NGC\,2516 are elongated, their elongation is less significant than that of Blanco\,1. All three clusters tend to stretch along the direction of the Galactic plane. Considering the age of $\sim$100\,Myr, such changes in shape are most likely caused by the external tidal field-stretching \citep{dinnbier2020}.

In Figure~\ref{fig:3d} panels (1.b--3.b), we display the spatial distributions of the clusters in the $X$-$Z$ plane. Despite the similarity in age, the morphology of stellar population outside the tidal radius differs significantly. Earlier studies have already discovered tidal tails in Blanco\,1 \citep{zhang20,pang2021}. 
37\% of the mass of Blanco\,1 is unbound and is located in the tidal tails.  For the more massive Pleiades and NGC 2516, however, 14\% of the mass (102.5\,$\rm M_{\odot}$ and 274.8\,$\rm M_{\odot}$, respectively) is located outside the tidal radii.
At a given Galactic position, the cluster mass is a dominant factor in shaping the 3D spatial morphology. Therefore, the more massive Pleiades and NGC\,2516 can withstand the external tidal stretching for a longer time. Their structures outside the tidal radius may indicate an early tidal disruption, while Blanco\,1 is already in an advanced stage of tidal disruption.

 To study the tidal structures in Pleiades and NGC\,2516, we convert the proper motions of members into tangential velocities (panels (1.c--3.c) in Figure~\ref{fig:3d}). All clusters are elongated along $V_{t_{\delta}}$, forming ``kinematic tails''. The bound stars within the tidal radius (grey circles) are concentrated in kinematic space, whose kinematics is dominated by the internal two-body relaxation and follows the cluster potential. On the other hand, the ``kinematic tails'' are mainly composed of unbound stars influenced by Galactic tide. 
We divide the kinematic tails into two subgroups: one with $V_{t_{\delta}}$ larger (red circles) than the median value (black crosses), and the other smaller than the median value (blue circles).
These two subgroups are distinct in the 3D spatial distribution (panels b), located on opposite sides of the cluster. In Blanco\,1, the red and blue kinematic tails correspond to the trailing and leading tails, which resembles the proper motion tails found by \citet{zhang20}. Likewise the structure outside the tidal radius in the Pleiades and NGC\,2516 are well separated into two parts by the red and blue kinematic tails. This correlation supports an early tidal disruption in these two clusters, with tidal tails just being formed.

The kinematic tails are most apparent in the Pleiades, with a velocity difference of 3\,km\,s$^{-1}$, as compared to a difference of 2\,km\,s$^{-1}$ in Blanco\,1 and NGC\,2516. This variation is a consequence of the different viewing angles. The elongated kinematic tails are induced by the Galactic differential rotation in a group of mainly unbound stars, spanning several tens to 100\,pc in size. The internal velocity dispersion also plays a role in broadening the kinematic distribution, e.g., the most massive NGC\,2516 has the highest velocity dispersion ($\sigma_{V_{t_{\alpha}}}$, $\sigma_{V_{t_{\delta}}}$).

\section{Conclusion}

The ``kinematic tails'' in the tangential velocity space successfully reveal that stars outside the tidal radius in the Pleiades and NGC\,2516 are early-stage tidal structures, compared to the profoundly disrupting Blanco\,1.

\begin{acknowledgments}
This study is supported by the Research Development Fund of Xi'an Jiaotong-Liverpool University (RDF-18--02--32) and the XJTLU Summer Undergraduate Research Fund (SURF-2021119).
\end{acknowledgments}



\bibliography{main}{}

\begin{thebibliography}{}
\expandafter\ifx\csname natexlab\endcsname\relax\def\natexlab#1{#1}\fi
\providecommand{\url}[1]{\href{#1}{#1}}
\providecommand{\dodoi}[1]{doi:~\href{http://doi.org/#1}{\nolinkurl{#1}}}
\providecommand{\doeprint}[1]{\href{http://ascl.net/#1}{\nolinkurl{http://ascl.net/#1}}}
\providecommand{\doarXiv}[1]{\href{https://arxiv.org/abs/#1}{\nolinkurl{https://arxiv.org/abs/#1}}}

\bibitem[{{Dinnbier} \& {Kroupa}(2020)}]{dinnbier2020}
{Dinnbier}, F., \& {Kroupa}, P. 2020, \aap, 640, A85,
  \dodoi{10.1051/0004-6361/201936572}

\bibitem[{{Gaia Collaboration} {et~al.}(2021){Gaia Collaboration}, {Brown},
  {Vallenari}, {Prusti}, {de Bruijne}, {Babusiaux}, {Biermann}, {Creevey},
  {Evans}, {Eyer}, {Hutton}, {Jansen}, {Jordi}, {Klioner}, {Lammers},
  {Lindegren}, {Luri}, {Mignard}, {Panem}, {Pourbaix}, {Randich}, {Sartoretti},
  {Soubiran}, {Walton}, {Arenou}, {Bailer-Jones}, {Bastian}, {Cropper},
  {Drimmel}, {Katz}, {Lattanzi}, {van Leeuwen}, {Bakker}, {Cacciari},
  {Casta{\~n}eda}, {De Angeli}, {Ducourant}, {Fabricius}, {Fouesneau},
  {Fr{\'e}mat}, {Guerra}, {Guerrier}, {Guiraud}, {Jean-Antoine Piccolo},
  {Masana}, {Messineo}, {Mowlavi}, {Nicolas}, {Nienartowicz}, {Pailler},
  {Panuzzo}, {Riclet}, {Roux}, {Seabroke}, {Sordo}, {Tanga}, {Th{\'e}venin},
  {Gracia-Abril}, {Portell}, {Teyssier}, {Altmann}, {Andrae}, {Bellas-Velidis},
  {Benson}, {Berthier}, {Blomme}, {Brugaletta}, {Burgess}, {Busso}, {Carry},
  {Cellino}, {Cheek}, {Clementini}, {Damerdji}, {Davidson}, {Delchambre},
  {Dell'Oro}, {Fern{\'a}ndez-Hern{\'a}ndez}, {Galluccio}, {Garc{\'\i}a-Lario},
  {Garcia-Reinaldos}, {Gonz{\'a}lez-N{\'u}{\~n}ez}, {Gosset}, {Haigron},
  {Halbwachs}, {Hambly}, {Harrison}, {Hatzidimitriou}, {Heiter},
  {Hern{\'a}ndez}, {Hestroffer}, {Hodgkin}, {Holl}, {Jan{\ss}en}, {Jevardat de
  Fombelle}, {Jordan}, {Krone-Martins}, {Lanzafame}, {L{\"o}ffler}, {Lorca},
  {Manteiga}, {Marchal}, {Marrese}, {Moitinho}, {Mora}, {Muinonen}, {Osborne},
  {Pancino}, {Pauwels}, {Petit}, {Recio-Blanco}, {Richards}, {Riello},
  {Rimoldini}, {Robin}, {Roegiers}, {Rybizki}, {Sarro}, {Siopis}, {Smith},
  {Sozzetti}, {Ulla}, {Utrilla}, {van Leeuwen}, {van Reeven}, {Abbas}, {Abreu
  Aramburu}, {Accart}, {Aerts}, {Aguado}, {Ajaj}, {Altavilla}, {{\'A}lvarez},
  {{\'A}lvarez Cid-Fuentes}, {Alves}, {Anderson}, {Anglada Varela}, {Antoja},
  {Audard}, {Baines}, {Baker}, {Balaguer-N{\'u}{\~n}ez}, {Balbinot}, {Balog},
  {Barache}, {Barbato}, {Barros}, {Barstow}, {Bartolom{\'e}}, {Bassilana},
  {Bauchet}, {Baudesson-Stella}, {Becciani}, {Bellazzini}, {Bernet}, {Bertone},
  {Bianchi}, {Blanco-Cuaresma}, {Boch}, {Bombrun}, {Bossini}, {Bouquillon},
  {Bragaglia}, {Bramante}, {Breedt}, {Bressan}, {Brouillet}, {Bucciarelli},
  {Burlacu}, {Busonero}, {Butkevich}, {Buzzi}, {Caffau}, {Cancelliere},
  {C{\'a}novas}, {Cantat-Gaudin}, {Carballo}, {Carlucci}, {Carnerero},
  {Carrasco}, {Casamiquela}, {Castellani}, {Castro-Ginard}, {Castro Sampol},
  {Chaoul}, {Charlot}, {Chemin}, {Chiavassa}, {Cioni}, {Comoretto}, {Cooper},
  {Cornez}, {Cowell}, {Crifo}, {Crosta}, {Crowley}, {Dafonte}, {Dapergolas},
  {David}, {David}, {de Laverny}, {De Luise}, {De March}, {De Ridder}, {de
  Souza}, {de Teodoro}, {de Torres}, {del Peloso}, {del Pozo}, {Delbo},
  {Delgado}, {Delgado}, {Delisle}, {Di Matteo}, {Diakite}, {Diener},
  {Distefano}, {Dolding}, {Eappachen}, {Edvardsson}, {Enke}, {Esquej}, {Fabre},
  {Fabrizio}, {Faigler}, {Fedorets}, {Fernique}, {Fienga}, {Figueras},
  {Fouron}, {Fragkoudi}, {Fraile}, {Franke}, {Gai}, {Garabato},
  {Garcia-Gutierrez}, {Garc{\'\i}a-Torres}, {Garofalo}, {Gavras}, {Gerlach},
  {Geyer}, {Giacobbe}, {Gilmore}, {Girona}, {Giuffrida}, {Gomel}, {Gomez},
  {Gonzalez-Santamaria}, {Gonz{\'a}lez-Vidal}, {Granvik},
  {Guti{\'e}rrez-S{\'a}nchez}, {Guy}, {Hauser}, {Haywood}, {Helmi}, {Hidalgo},
  {Hilger}, {H{\l}adczuk}, {Hobbs}, {Holland}, {Huckle}, {Jasniewicz},
  {Jonker}, {Juaristi Campillo}, {Julbe}, {Karbevska}, {Kervella}, {Khanna},
  {Kochoska}, {Kontizas}, {Kordopatis}, {Korn}, {Kostrzewa-Rutkowska},
  {Kruszy{\'n}ska}, {Lambert}, {Lanza}, {Lasne}, {Le Campion}, {Le Fustec},
  {Lebreton}, {Lebzelter}, {Leccia}, {Leclerc}, {Lecoeur-Taibi}, {Liao},
  {Licata}, {Lindstr{\o}m}, {Lister}, {Livanou}, {Lobel}, {Madrero Pardo},
  {Managau}, {Mann}, {Marchant}, {Marconi}, {Marcos Santos}, {Marinoni},
  {Marocco}, {Marshall}, {Martin Polo}, {Mart{\'\i}n-Fleitas}, {Masip},
  {Massari}, {Mastrobuono-Battisti}, {Mazeh}, {McMillan}, {Messina},
  {Michalik}, {Millar}, {Mints}, {Molina}, {Molinaro}, {Moln{\'a}r},
  {Montegriffo}, {Mor}, {Morbidelli}, {Morel}, {Morris}, {Mulone}, {Munoz},
  {Muraveva}, {Murphy}, {Musella}, {Noval}, {Ord{\'e}novic}, {Orr{\`u}},
  {Osinde}, {Pagani}, {Pagano}, {Palaversa}, {Palicio}, {Panahi}, {Pawlak},
  {Pe{\~n}alosa Esteller}, {Penttil{\"a}}, {Piersimoni}, {Pineau}, {Plachy},
  {Plum}, {Poggio}, {Poretti}, {Poujoulet}, {Pr{\v{s}}a}, {Pulone}, {Racero},
  {Ragaini}, {Rainer}, {Raiteri}, {Rambaux}, {Ramos}, {Ramos-Lerate}, {Re
  Fiorentin}, {Regibo}, {Reyl{\'e}}, {Ripepi}, {Riva}, {Rixon}, {Robichon},
  {Robin}, {Roelens}, {Rohrbasser}, {Romero-G{\'o}mez}, {Rowell}, {Royer},
  {Rybicki}, {Sadowski}, {Sagrist{\`a} Sell{\'e}s}, {Sahlmann}, {Salgado},
  {Salguero}, {Samaras}, {Sanchez Gimenez}, {Sanna}, {Santove{\~n}a},
  {Sarasso}, {Schultheis}, {Sciacca}, {Segol}, {Segovia}, {S{\'e}gransan},
  {Semeux}, {Shahaf}, {Siddiqui}, {Siebert}, {Siltala}, {Slezak}, {Smart},
  {Solano}, {Solitro}, {Souami}, {Souchay}, {Spagna}, {Spoto}, {Steele},
  {Steidelm{\"u}ller}, {Stephenson}, {S{\"u}veges}, {Szabados}, {Szegedi-Elek},
  {Taris}, {Tauran}, {Taylor}, {Teixeira}, {Thuillot}, {Tonello}, {Torra},
  {Torra}, {Turon}, {Unger}, {Vaillant}, {van Dillen}, {Vanel}, {Vecchiato},
  {Viala}, {Vicente}, {Voutsinas}, {Weiler}, {Wevers}, {Wyrzykowski}, {Yoldas},
  {Yvard}, {Zhao}, {Zorec}, {Zucker}, {Zurbach}, \& {Zwitter}}]{gaia21}
{Gaia Collaboration}, {Brown}, A.~G.~A., {Vallenari}, A., {et~al.} 2021, \aap,
  649, A1, \dodoi{10.1051/0004-6361/202039657}

\bibitem[{{Lodieu} {et~al.}(2019){Lodieu}, {P{\'e}rez-Garrido}, {Smart}, \&
  {Silvotti}}]{Lodieu19}
{Lodieu}, N., {P{\'e}rez-Garrido}, A., {Smart}, R.~L., \& {Silvotti}, R. 2019,
  \aap, 628, A66, \dodoi{10.1051/0004-6361/201935533}

\bibitem[{{Pang} {et~al.}(2021){Pang}, {Li}, {Yu}, {Tang}, {Dinnbier},
  {Kroupa}, {Pasquato}, \& {Kouwenhoven}}]{pang2021}
{Pang}, X., {Li}, Y., {Yu}, Z., {et~al.} 2021, \apj, 912, 162,
  \dodoi{10.3847/1538-4357/abeaac}

\bibitem[{{R{\"o}ser} \& {Schilbach}(2019)}]{roser2019b}
{R{\"o}ser}, S., \& {Schilbach}, E. 2019, \aap, 627, A4,
  \dodoi{10.1051/0004-6361/201935502}

\bibitem[{{R{\"o}ser} {et~al.}(2019){R{\"o}ser}, {Schilbach}, \&
  {Goldman}}]{roser2019a}
{R{\"o}ser}, S., {Schilbach}, E., \& {Goldman}, B. 2019, \aap, 621, L2,
  \dodoi{10.1051/0004-6361/201834608}

\bibitem[{{Tang} {et~al.}(2019){Tang}, {Pang}, {Yuan}, {Chen}, {Hong},
  {Goldman}, {Just}, {Shukirgaliyev}, \& {Lin}}]{tan19}
{Tang}, S.-Y., {Pang}, X., {Yuan}, Z., {et~al.} 2019, \apj, 877, 12,
  \dodoi{10.3847/1538-4357/ab13b0}

\bibitem[{{Yuan} {et~al.}(2018){Yuan}, {Chang}, {Banerjee}, {Han}, {Kang}, \&
  {Smith}}]{yuan18}
{Yuan}, Z., {Chang}, J., {Banerjee}, P., {et~al.} 2018, \apj, 863, 26,
  \dodoi{10.3847/1538-4357/aacd0d}

\bibitem[{{Zhang} {et~al.}(2020){Zhang}, {Tang}, {Chen}, {Pang}, \&
  {Liu}}]{zhang20}
{Zhang}, Y., {Tang}, S.-Y., {Chen}, W.~P., {Pang}, X., \& {Liu}, J.~Z. 2020,
  \apj, 889, 99, \dodoi{10.3847/1538-4357/ab63d4}

\end{thebibliography}
\bibliographystyle{aasjournal}


\end{document}